\documentstyle[seceq,epsf]{ptptex}

\begin{document}

\title{Monte Carlo simulation of GaAs(001) homoepitaxy}
\author{M.\ Itoh$^{1,}$\footnote{
Present address: Department of Physics, Graduate School of Science,
Osaka University, Toyonaka, Osaka, 560-0043 Japan.
E-mail address: makoto@acty.phys.sci.osaka-u.ac.jp}, 
G.\ R.\ Bell$^{2}$, B.\ A.\ Joyce$^{1}$, and D.\ D.\ Vvedensky$^{1}$}
\inst{
$^{1}$Department of Physics, Imperial College, London SW7 2BZ, UK. \\
$^{2}$Centre for Electronic Materials and Devices, Imperial College, 
London SW7 2AB, UK.}
\maketitle
\begin{abstract}
By carrying out Monte Carlo simulations based on the two-species atomic-scale 
kinetic growth model of GaAs(001) homoepitaxy and comparing the results with 
scanning tunneling microscope images, we show that initial growing 
islands undergo the structural transformation before adopting the proper 
$\beta2(2\times4)$ reconstruction. 
\end{abstract}

\begin{figure}
	\epsfxsize = 6 cm
	\centerline{\epsfbox{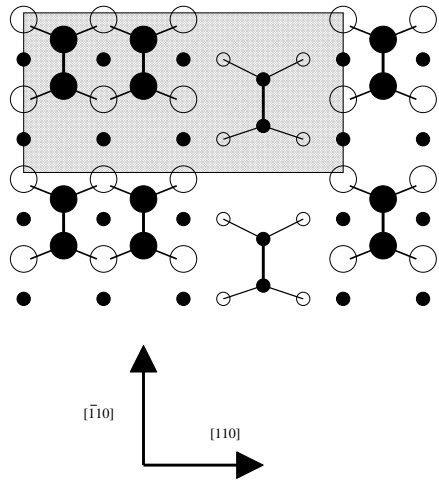}}
	\caption{GaAs(001)-$\beta2(2\times4)$ structure.
	The dark and bright disks denote, respectively, As and Ga atoms, 
	whose radii decrease according to their depths.  
	The shaded area denotes the unit cell.}
	\label{fig:2x4}
\end{figure}

In developing optoelectronic devices, GaAs(001) is often used as the basis 
substrate for fabrication~\cite{kelly95}. 
To study the atomic-scale growth kinetics of this surface, extensive use 
has been made of {\it ab initio} calculations 
with the energy-minimization procedure~\cite{kratzer99A}. 
With this method, however, it is actually difficult to elucidate the true 
growth kinetics right at the growth temperatures, 
because growth proceeds in non-equilibrium conditions there. 
For example, by investigating the stable sites for Ga adatoms and assuming 
arsenic species to stick onto them, it is deduced in Ref.~\cite{kratzer99A} 
that the nucleation of islands is initiated by the sticking of Ga adatoms at 
the trench sites of the GaAs(001)-$\beta2(2\times4)$ structure, which is 
depicted in Fig.\ref {fig:2x4}. 
If this growth mechanism is true, a quite large portion of a GaAs(001) 
surface must reduce the widths of the trenches from double to single before 
islands nucleate. 
However, no scanning tunneling microscope (STM) images obtained thus far 
seem to support it. 

Here we carry out atomic-scale Monte Carlo (MC) simulations as the alternative 
means to investigate nucleation and growth of islands on 
the GaAs(001)-$\beta2(2\times4)$ surface. 
For this purpose, we developed the two-species model of homoepitaxial 
growth on the GaAs(001)-$\beta2(2\times4)$ structure~\cite{itoh98B}. 

\begin{figure}
 \epsfxsize = 6 cm
 \centerline{\epsfbox{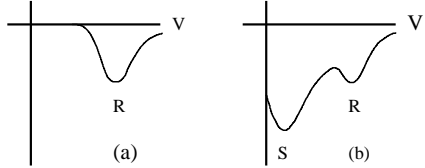}}
 \caption{Incorporation kinetics of As$_2$ molecules. 
 (a) Ga atoms are not properly arranged on a surface, so that the 
 incorporation is prohibited. Only the desorption of As$_2$ 
 (${\rm R} \rightarrow {\rm V}$) takes place. 
 (b) Incorporation (${\rm R} \rightarrow {\rm S}$) is enabled.
 The barrier we used for the ${\rm R} \rightarrow {\rm S}$ process is 1.74 eV, 
 and the barrier difference between the two processes 
 ${\rm R} \rightarrow {\rm S}$ and ${\rm R} \rightarrow {\rm V}$ is 0.65 eV. 
 } 
 \label{fig:potential}
\end{figure}
For the arsenic deposition process, we use As$_2$ molecules, for which we 
simplify the incorporation process onto a surface structure in a mean field 
manner by introducing the chemical reservoir, where deposited As$_2$ 
molecules are once stored. 
This process is schematically drawn in Figs.\ref{fig:potential} (a) and (b), 
where R, S, and V denote, respectively, the reservoir state, 
the surface-incorporated state, and the vacuum, which is far distant from 
a surface. 

As Figs.\ref{fig:potential} (a) and (b) indicate, whether an As$_2$ molecule 
can be incorporated onto a surface site depends on if surface Ga atoms are 
properly arranged there or not. 
When they are properly arranged, the incorporation channel of an As$_2$ 
molecules becomes open, and the potential for As$_2$ molecules changes from 
Fig.\ref{fig:potential}(a) to Fig.\ref{fig:potential}(b). 
As a consequence, the macroscopic growth rate is controlled by the deposition 
rate of Ga atoms. 
However, at atomic scales, arsenic species come into play in the growth 
kinetics, as we will see below. 
Note that this treatment of the chemisorption process improves the efficiency 
of computation, so that we can study the morphology of a growing surface 
in a realistic computation time. 

\begin{figure}
 \epsfxsize = 6 cm
 \centerline{\epsfbox{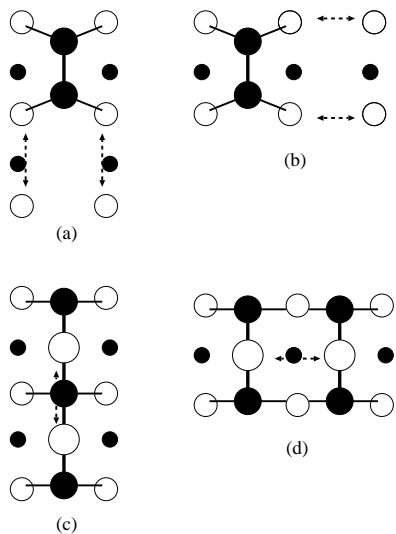}}
 \caption{Ga-part of the interactions. 
 The interaction energies are (in eV):~(a)$0.28$, (b)$0.25$, (c)$0.28$, 
 (d)$0.32$, and the back-bonding energy is $1.40\;$eV per atom.  
 }\label{Ga}
\end{figure}
\begin{figure}
 \epsfxsize = 6 cm
 \centerline{\epsfbox{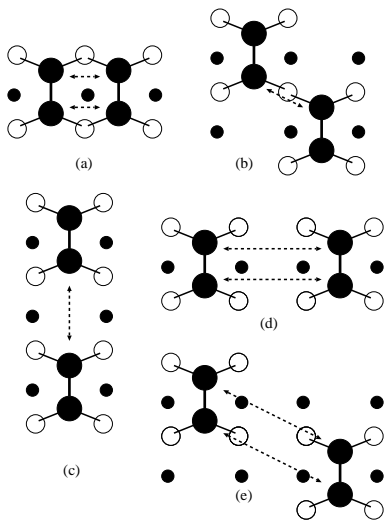}}
 \caption{As-part of the interactions. 
 The interaction energies are (in eV):~(a)$1.00$, (b)$0.63$, 
 (c)$-0.15$, (d)$-1.21$, (e)$-0.25$. 
 The back-bonding energy is $1.65\;$eV per atom. 
 }\label{As}
\end{figure}

For the interactions between surface atoms, we take account of those 
between the same species in the same layers, as depicted for the Ga-Ga 
interactions in Fig.\ref{Ga} and for the As-As interactions in Fig.\ref{As}. 

Among them, the interactions in Figs.\ref{As} (d) and (e), representing the 
Coulomb repulsions between two surface As atoms, play the central role in 
realizing the $\beta2(2\times4)$ structure~\cite{northrup94}. 
We note that these interaction parameters were determined solely by the 
comparison between the calculation results and the STM images. 

\begin{figure}
 \epsfxsize = 10.5 cm
 \centerline{\epsfbox{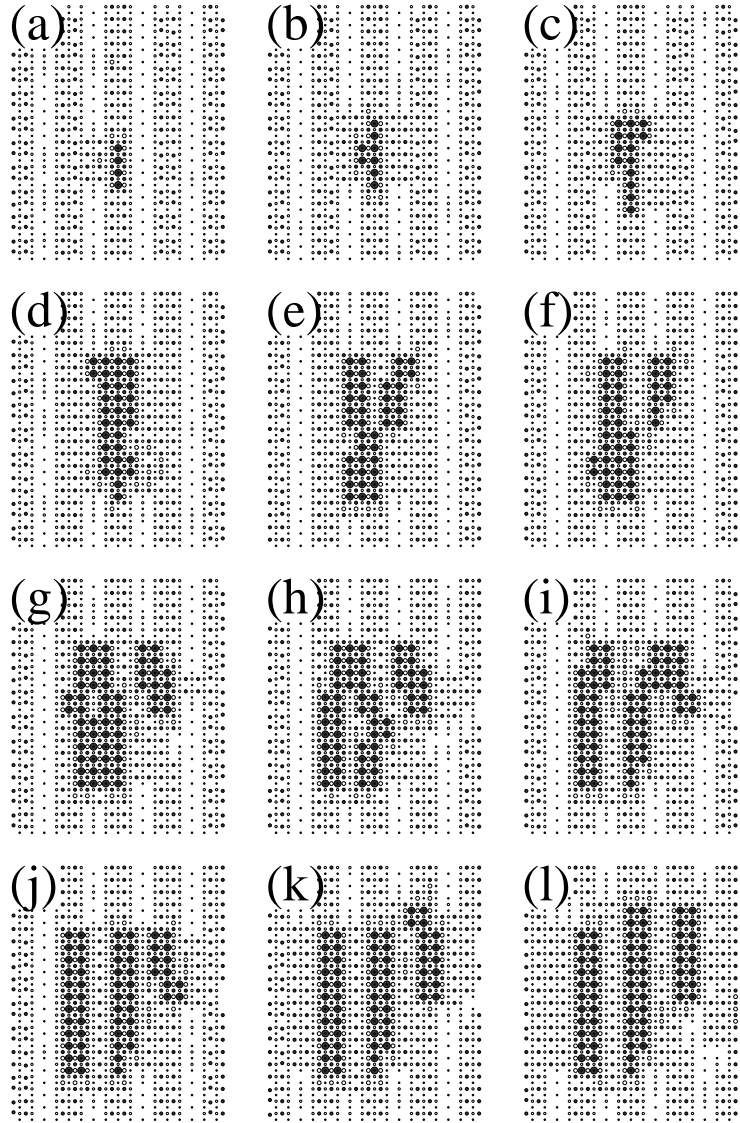}}
 \caption{Nucleation and growth of an island.}
 \label{fig:snapshot}
\end{figure}

By carrying out simulations with the kinetic MC algorithm~\cite{bortz}, 
we found that the nucleation of islands take place on top of 
the $\beta2(2\times4)$ substrate without having the $(2\times4)$ structure, 
which transform into the $(2\times4)$ structure at a later stage of 
growth~\cite{itoh98B}. 
This is because the excess charge on surface As atoms, which is the origin of 
the Coulomb repulsions in Figs.\ref{As}(d) and (e), is not significant 
on small islands. 
The processes of nucleation and the structural transformation are well 
exemplified by the series of the snapshots in Fig.\ref{fig:snapshot}. 

\begin{figure}
 \epsfxsize = 11 cm
 \centerline{\epsfbox{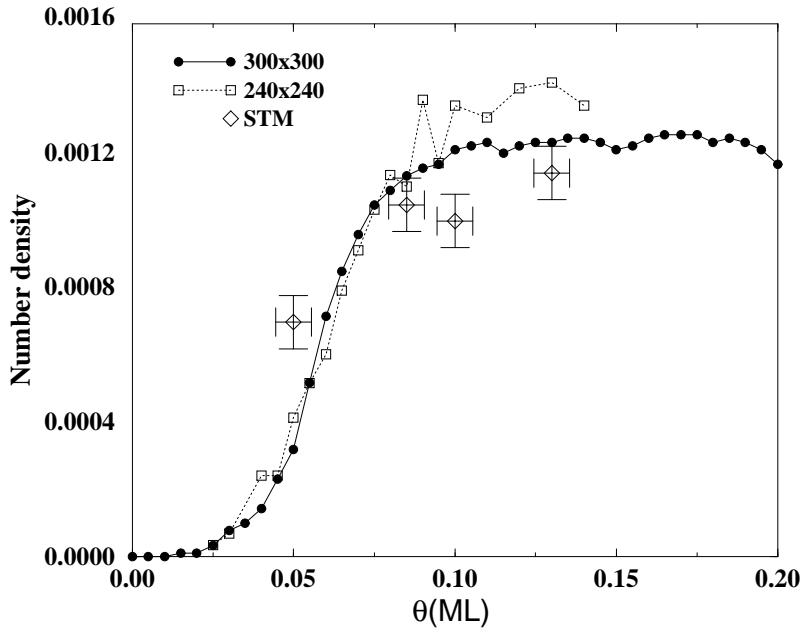}}
 \caption{Total number densities of islands as functions of a coverage 
 compared with STM measurements at the indicated coverages.}
 \label{fig:size}
\end{figure}

\begin{figure}
 \epsfxsize = 11 cm
 \centerline{\epsfbox{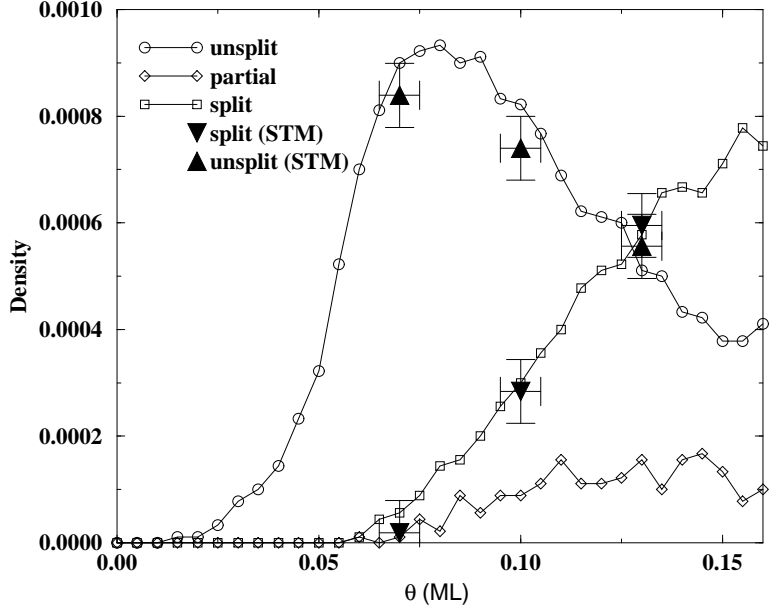}}
 \caption{Densities of split, unsplit, and partially split islands 
 obtained by simulations on a $300\times300$ lattice
 and STM observations.}
 \label{fig:types}
\end{figure}

To further confirm the validity of our model quantitatively, we plotted 
in Fig.\ref{fig:size} 
the island-size distributions obtained by carrying out the calculations on 
the $300\times300$ lattice as well as on the $240\times240$ lattice and 
compared them with the STM images obtained at several different coverages. 
Here the lattice spacing in the model calculations corresponds to 4.0\AA~ 
on a real surface. 
The result of the comparison for the total number of islands is displayed 
in Fig.\ref{fig:size}. 
In this comparison, the results of the calculations were obtained 
at $T = 580^{\circ}$C, whereas the STM images were obtained after quenching 
the surface down to room temperature. 
In order to avoid the ambiguity caused by this difference in the conditions, 
we examined the effect of quenching, i.e., 
we performed the quenching simulations to find that single As dimers appearing 
at $T = 580^{\circ}$C change into double ones by quenching, so that the number 
of islands does not change during this process. 
We further found that other transient non-$(2\times4)$ structures 
tend to adopt the $\beta2(2\times4)$ structure during quenching. 
These results indicate that we can avoid the ambiguity and, hence, we can 
carry out a reliable comparison between the simulations and the STM images 
if we count the numbers of islands before, during, and after the splitting 
separately by classifying them into unsplit, partially split, and split 
islands, and discard the partially split ones, because they appear only as 
transient non-$(2\times4)$ structures. 
The results are seen in Fig.\ref{fig:types} to show a very good agreement 
between them. 

Since Ga atoms prefer to reside at trench sites, As$_2$ molecules can stick 
onto them and the width of the trench does change locally from double to 
single, as suggested by the {\it ab initio} calculations~\cite{kratzer99A}. 
However, according to our calculations, these surface As dimers desorb from 
such sites within 1 ms at $T = 580^{\circ}$C, so that actually these processes 
do not result in the growth of islands. 
Instead, when Ga atoms gather on top of the As-dimer row of the 
$\beta2(2\times4)$ structure and an As$_2$ dimer sticks onto them, this 
is stable enough to be the nucleation site. 
Note that the complexity of this process is the origin of the significant 
delay in the nucleation time seen in Figs.\ref{fig:size} and \ref{fig:types}.

Our results show that the nucleation of an island is controlled by Ga
kinetics, whereas the structural transformation and, accordingly, the
appearance of the $\beta2(2\times4)$ structure on newly created islands
is due to the splitting process seen in Figs.~\ref{fig:snapshot} (d--l),
which are controlled principally by the Coulomb repulsions in
Figs.~\ref{As} (d) and (e), and thus, by As kinetics. 
By further employing the rate equations, we confirmed that these two are
the only rate-limiting processes at the pre-coalescence regime of
GaAs(001) homoepitaxy~\cite{itoh99}. 

In summary, by carrying out atomic-scale growth simulations of 
GaAs(001)-$\beta2(2\times4)$ structure, we found that the nucleation of 
islands takes place on top of the $(2\times4)$ substrate and the islands grow 
via the structural transformation from non-$(2\times4)$ structure 
to the $\beta2(2\times4)$ structure. 
We further showed the validity of our model study quantitatively by 
comparing the results with 
those obtained by scanning tunneling microscopy observation.

\acknowledgements
One of the authors (M.I.) would like to thank 
Dr.\ T.\ Ohno for his encouragement. 



\end{document}